\documentclass[onecolumn,superscriptaddress,nofootinbib,notitlepage,longbibliography,aps]{revtex4-2}
\pdfoutput=1

\usepackage{graphicx}
\usepackage{amsmath}
\usepackage{amssymb}
\usepackage{amsfonts}
\usepackage[dvipsnames]{xcolor}
\usepackage[linktoc=none]{hyperref}
\usepackage[utf8]{inputenc}
\usepackage{booktabs}
\usepackage{comment}
\usepackage{multirow}
\definecolor{tab-blue}{RGB}{0, 107, 164}
\hypersetup{colorlinks=true,allcolors=tab-blue}

\newcommand{\INFN}{INFN - Sezione di Napoli, Complesso Universitario Monte Sant'Angelo, 80126 Napoli, Italy}
\newcommand{\SSM}{Scuola Superiore Meridionale, Via Mezzocannone 4, 80138 Napoli, Italy}
\begin{document}

\title{Probing super-heavy dark
matter with ultra-high-energy gamma rays}

\author{Marco Chianese}
\email{m.chianese@ssmeridionale.it}
\affiliation{\SSM}
\affiliation{\INFN}
\author{Ninetta Saviano}
\email{nsaviano@na.infn.it}
\affiliation{\INFN}
\affiliation{\SSM}
\author{Sara Cesare}
%\email{s.cesare@ssmeridionale.it}
\affiliation{\SSM}
\affiliation{\INFN}
\author{Vincenzo M. Grieco}
%\email{v.grieco@ssmeridionale.it}
\affiliation{\SSM}
\affiliation{\INFN}
\author{Valentina Nasti}
%\email{v.nasti@ssmeridionale.it}
\affiliation{\SSM}
\affiliation{\INFN}
\author{Francesca Spinnato}
%\email{f.spinnato@ssmeridionale.it}
\affiliation{\SSM}
\affiliation{\INFN}
\author{Alessandro Tiano}
%\email{a.tiano@ssmeridionale.it}
\affiliation{\SSM}
\affiliation{\INFN}

\begin{abstract}
We refine the constraints on the lifetime of decaying super-heavy dark matter  particles  (SHDM), with masses ranging from $10^7$ to $10^{15}$~GeV, by analyzing ultra-high-energy (UHE) gamma-ray data. Our approach involves an accurate comparison of the primary gamma-ray emissions resulting from prompt SHDM decays in the galactic halo with the most recent upper limits on isotropic UHE gamma-ray fluxes provided by various extensive air shower  experiments. We demonstrate that a precise consideration of the field of view and the geometric acceptance of different UHE gamma-ray observatories has significant implications for the inferred limits of dark matter lifetime. In addition, we examine the influence of uncertainties linked to the current models of the galactic dark matter distribution, employing diverse halo density profiles while varying both their radial  extent and the local dark matter density. Our findings indicate that the newly established UHE gamma-ray constraints are marginally less stringent than earlier evaluations, thereby revisiting the SHDM parameter space and allowing for observable neutrino fluxes.
\end{abstract}

\maketitle
\tableofcontents

%========================================
\section{Introduction}
%========================================

Within the $\Lambda$CDM paradigm, approximately $26\%$ of the Universe’s matter-energy budget consists of a nonrelativistic, pressureless component, referred to as dark matter (DM), which interacts almost exclusively through gravity.  ~\cite{Bertone:2004pz, Bertone:2016nfn, Kahlhoefer:2017dnp, Billard:2021uyg, AlvesBatista:2021eeu}. Among the various strategies to probe DM particles, indirect detection searches play a central role by seeking the Standard Model (SM) byproducts of DM annihilation or decay in astrophysical environments. Motivated by the absence of clear signals from DM particles in the GeV--TeV mass range, increasing attention has shifted toward scenarios with super-heavy dark matter (SHDM) that can decay on cosmological timescales~\cite{Chang:1996vw, Greene:1997ge, Chung:1998zb, Benakli:1998ut, Kolb:1998ki, Faraggi:1999iu, Coriano:2001mg, Aloisio:2006yi, Aloisio:2007bh, Chianese:2016smc, Kannike:2016jfs, Kolb:2017jvz, Kim:2019udq, Dudas:2020sbq, Hambye:2020lvy, Ling:2021zlj, Allahverdi:2023nov}. In such models, the dark matter lifetime is much longer than the age of the Universe, ensuring consistency with cosmological and astrophysical observations, while rare decay events can still produce observable fluxes of SM particles, notably cosmic rays, high-energy neutrinos, and Ultra-High-Energy (UHE) gamma rays~\cite{Murase:2012xs, Esmaili:2012us, Esmaili:2013gha, Rott:2014kfa, Boucenna:2015tra, Esmaili:2015xpa, Murase:2015gea, Cohen:2016uyg, Kalashev:2016cre, DelleRose:2017vvz, IceCube:2018tkk, Kachelriess:2018rty, Chianese:2019kyl, Dekker:2019gpe, Ishiwata:2019aet, Chianese:2021htv, Esmaili:2021yaw,  Maity:2021umk, Guepin:2021ljb, Chianese:2021jke, Arguelles:2022nbl, LHAASO:2022yxw, Skrzypek:2022hpy, PierreAuger:2022jyk, PierreAuger:2022ubv, Das:2023wtk, Fiorillo:2023clw, Leung:2023gwp, PierreAuger:2023vql, Munbodh:2024ast, Sarmah:2024ffy, Berghaus:2025jwb, Boehm:2025qro, Dubey:2025ouh, Rocamora:2025ddt}. Interestingly, decaying SHDM particles have been proposed as a possible explanation of the recent KM3NeT/ARCA observation of a neutrino event with
an energy of approximately 220 PeV~\cite{KM3NeT:2025npi, Barman:2025hoz, Jho:2025gaf, Aloisio:2025nts, Kohri:2025bsn, Borah:2025igh, Khan:2025gxs, Murase:2025uwv}. Hence, it is crucial to assess the viability of this scenario by robustly quantifying the allowed SHDM parameter space from a multi-messenger perspective.

In this work, we examine the prompt gamma-ray emission of decaying SHDM particles with masses from $10^7$ to $10^{15}$~GeV, originating from the galactic DM halo, for several decay channels. Hence, we derive the constraints on the SHDM lifetime by exploiting the latest upper bounds on the isotropic flux of UHE gamma rays placed by the extensive air shower (EAS) experiments CASA-MIA~\cite{CASA-MIA:1997tns}, KASCADE~\cite{KASCADE:2005ynk}, KASCADE-Grande~\cite{KASCADEGrande:2017vwf}, Telescope Array (TA)~\cite{TelescopeArray:2018rbt} and Pierre Auger Observatory (PAO)~\cite{PierreAuger:2025jwt, PierreAuger:2022uwd, PierreAuger:2024ayl, PierreAuger:2022aty}. Not only do these observatories probe the UHE gamma-ray sky over complementary energy ranges, but also have distinct sky coverages and sensitivities, owing to their different latitudes and zenith-angle acceptances. We improve upon previous SHDM analyses by explicitly accounting for the field of view (FOV) and the geometrical acceptance efficiency of each experiment. As a result, the $\mathcal{D}$-factor, which represents the line-of-sight integral of the DM density, varies between experiments, thus influencing the DM constraints. Moreover, we adopt updated parameter values for two representative galactic DM halo models, namely the generalized Navarro-Frenk-White (gNFW) and the Burkert density profiles, and assess the robustness of the resulting SHDM lifetime constraints with respect to the uncertainties on determination of the galactic DM distribution~\cite{Benito:2019ngh, Benito:2020lgu}.

The paper is organized as follows.
\textcolor{black}{In Section~\ref{sec:flux}, we present a detailed calculation of the UHE gamma-ray flux produced by decaying SHDM particles, with particular emphasis on the evaluation of the energy-dependent $\mathcal{D}$-factor for both the DM density profiles.}
In Section~\ref{sec:constraints}, we outline the statistical methodology and report the resulting lower bounds on the SHDM lifetime derived from the latest UHE gamma-ray data. Finally, in Section~\ref{sec:conclusions}, we draw the main conclusions of this work.

%========================================
\section{Gamma-ray flux from heavy decaying dark matter}
\label{sec:flux}
%========================================

A possible origin of UHE gamma rays is the decay of very massive DM particles. The overall gamma-ray signal expected from DM includes several components, arising from both prompt and secondary radiation produced in the Galaxy and beyond. In this study, we restrict ourselves to the prompt galactic emission, which dominates the spectrum at the highest photon energies reachable in DM-decay scenarios, while omitting the extragalactic contribution because it is heavily attenuated at UHE energies through $\gamma\gamma$ interactions. Furthermore,  we also disregard the secondary galactic emission  from Inverse Compton scattering of electrons and positrons on low-energy photon backgrounds (see, e.g., Ref.~\cite{Ishiwata:2019aet}). Indeed,  this component is typically subleading compared with the prompt galactic one, and we have verified that its inclusion would not significantly impact our conclusions. This simplified approach is adequate for deriving strong and comprehensive gamma-ray constraints on dark matter, using observations that span several decades in energy, ranging from $10^{5}$ up to $10^{11}$~GeV.

The UHE gamma-ray flux from the prompt galactic emission of DM decays is expressed by
\begin{equation}\label{eq:flux}
    \frac{\mathrm{d}\Phi_\gamma}{\mathrm{d}E_\gamma\mathrm{d}\Omega}(E_\gamma,b,l)=\frac{1}{4\pi\, m_\mathrm{DM}\tau_\mathrm{DM}}\frac{\mathrm{d}N_\gamma}{\mathrm{d}E_\gamma}\int_0^\infty \rho_\mathrm{DM}\left(r(s,b,l)\right)\,e^{-\tau_{\gamma\gamma}(E_\gamma,s,b,l)}\,\mathrm{d}s\,.
\end{equation}
Here, $m_{\rm DM}$ and $\tau_{\rm DM}$ respectively represent the mass and lifetime of the DM particle and purely depend on the nature of the DM candidate. The term ${\mathrm{d}N_\gamma}/{\mathrm{d}E_\gamma}(E_\gamma)$ is the energy spectrum of the generated photons and depends on the specific DM decay channel, which we compute using the \texttt{HDMSpectra} code~\cite{Bauer:2020jay} assuming a 100\% branching ratio at a time. The integral term is related to astrophysical quantities through the galactic DM density distribution $\rho_{\rm DM}$ and the transmission factor (with $\tau_{\gamma\gamma}$ being the optical depth) due to high-energy photon interactions with the photon fields in the galaxy. These quantities depend on the line-of-sight $s$, the galactic angular coordinates ($b$, $l$), and the galactocentric radial distance $r$ defined as
\begin{equation}\label{eq:rcoord}
    r(s,b,l) = \sqrt{s^2+R_\odot^2-2sR_\odot\cos b\cos l}~,
\end{equation}
where we fix the Sun’s
galactocentric distance to be $R_\odot = 8.178~\mathrm{kpc}$ according to the analyses in Refs.~\cite{Gravity:2019nxk, Benito:2020lgu}.

\textcolor{black}{We evaluate the transmission factor in Eq.~\eqref{eq:flux} following Refs.~\cite{Esmaili:2015xpa, Chianese:2021jke}. The optical depth $\tau_{\gamma\gamma}$ receives two contributions: one from the Cosmic Microwave Background (CMB) and one from the galactic starlight and infrared (SL+IR) radiation. The CMB component is isotropic and spatially uniform, and dominates the absorption in the range $10^6$--$10^7$~GeV, peaking at $E_\gamma \simeq 2\times10^6$~GeV. The SL+IR component, instead, traces the galactic distribution of dust and stars and introduces a non-trivial dependence on the line-of-sight direction, with a strong enhancement toward the galactic center and disk and with a rapid fall-off with latitude. Its higher target-photon energies lower the pair-production threshold, yielding a secondary absorption feature at $E_\gamma \simeq 10^5$~GeV~\cite{Moskalenko:2006}. To compute the SL+IR contribution to the optical depth, we take the corresponding photon density from the \texttt{GALPROP} code~\cite{galprop} (see also Refs.~\cite{Vernetto:2016alq, Porter:2017vaa, Popescu:2017dgo}).}
\begin{table}[t!]
    \centering
    \hspace*{-0.5em} 
    \begin{tabular}{l @{\hspace{1.5em}} c @{\hspace{1.5em}} c @{\hspace{1.5em}} c} 
        \toprule 
        \textbf{Experiment} & \textbf{Latitude ($^{\circ}$)} & \textbf{Zenith angle range ($^{\circ}$)} & \textbf{Energy range (GeV)} \\
        \midrule 
        CASA-MIA~ \cite{CASA-MIA:1997tns} & $~40.2$ & $~0-60$ & $3.26 \times 10^{5} - 3.33 \times 10^{7~}$ \\
        KASCADE~\cite{KASCADE:2005ynk} & $~49.1$ & $~0-18$ & $2.45 \times 10^{5} - 1.84 \times 10^{7~}$ \\
        KASCADE-Grande~\cite{KASCADEGrande:2017vwf} & $~49.1$ & $~0 - 40$ & $1.50 \times 10^{6} - 2.92 \times 10^{8~}$ \\
        TA~\cite{TelescopeArray:2018rbt} & $~39.3$ & $~0-55$ & $1.00 \times 10^{9} - 9.92 \times 10^{10}$ \\
        PAO UMD + SD 433 m~\cite{PierreAuger:2025jwt} & $-35.2~$ & $~0-52$ & $5.00 \times 10^{7} -2.00 \times 10^{8~}$ \\
        PAO HECO + SD 750 m~\cite{PierreAuger:2022uwd} & $-35.2~$ &  $~0-60$ & $2.00 \times 10^{8} - 1.00 \times 10^{9~}$ \\
        PAO Hybrid~\cite{PierreAuger:2024ayl} & $-35.2~$ & $~0-60$ & $1.00 \times 10^{9} - 9.96 \times 10^{9~}$ \\
        PAO SD 1500 m~\cite{PierreAuger:2022aty} & $-35.2~$ & $30-60$ & $9.96 \times 10^{9} - 3.98 \times 10^{10}$ \\
        \midrule
        %\bottomrule 
    \end{tabular}
    \hspace*{-0.5em} 
    \caption{Summary of the key parameters of the EAS experiments considered in this work, including latitude, zenith angle acceptance and energy range.}
\label{tab:experiments_parameters}
\end{table}
\begin{figure}[t!]
    \centering
    \includegraphics[width=0.5 \linewidth]{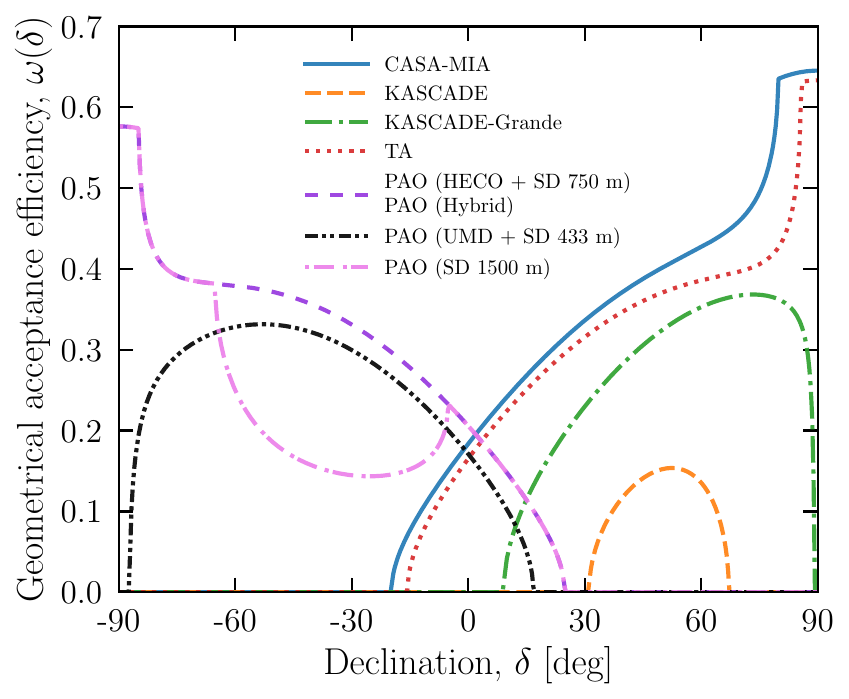}
    \caption{Geometrical acceptance efficiency $\omega(\delta)$ as a function of the declination $\delta$, for different UHE gamma-ray detectors shown with distinct curve styles. The configurations HECO + SD 750 m and Hybrid of the PAO detector share the same geometrical acceptance efficiency as their zenith angle ranges are equal (see Tab.~\ref{tab:experiments_parameters}).}
    \label{fig:efficiency_curves}
\end{figure}

To enable a direct comparison with existing bounds on the diffuse UHE gamma-ray flux, from Eq.~\eqref{eq:flux} we evaluate the angle-averaged integral photon flux, which for each experiment is defined as
\begin{equation}\label{eq:integral}
    \Phi_\gamma(E_\gamma)
    = \frac{1}{\Delta \Omega_{\rm exp}}
      \int_{E_\gamma}^{\infty}\mathrm{d}E^\prime_\gamma
      \int_{\Delta \Omega_{\rm exp}}
      \frac{\mathrm{d}\Phi_\gamma}{\mathrm{d}E^\prime_\gamma\,\mathrm{d}\Omega} \,\mathrm{d}\Omega = \frac{1}{4\pi\, m_\mathrm{DM}\tau_\mathrm{DM}}\int_{E_\gamma}^{\infty}\mathcal{D}_{\rm exp}(E^\prime_\gamma)\left.\frac{\mathrm{d}N_\gamma}{\mathrm{d}E_\gamma}\right|_{E^\prime_\gamma}\mathrm{d}E^\prime_\gamma\,,
\end{equation}
where $\Delta \Omega_{\rm exp}$ is the FOV of the experiment, and $\mathcal{D}_{\rm exp}(E_\gamma)$ is the energy-dependent $\mathcal{D}$-factor defined as
\begin{equation}\label{eq:Dfactor}
    \mathcal{D}_{\rm exp}(E_\gamma) = \frac{1}{\Delta \Omega_{\rm exp}}\int_{\Delta \Omega_{\rm exp}} \mathrm{d}\Omega \, \int_0^\infty \rho_\mathrm{DM}\left(r(s,b,l)\right)\,e^{-\tau_{\gamma\gamma}(E_\gamma,s,b,l)}\,\omega_{\rm exp}\left(\delta(b,l)\right)\,\mathrm{d}s\,,
\end{equation}
\textcolor{black}{with $\omega_{\rm exp}(\delta)$ being the geometrical acceptance efficiency of the experiment, which takes the expression (see Appendix~\ref{sec:efficiency} for more details on its derivation)
\begin{equation}\label{eq:omega_delta}
    \omega_{\rm exp}(\delta) = \frac{1}{\pi}\left(\alpha_m \sin\lambda\sin\delta + \cos\lambda \cos\delta \sin\alpha_m\right)\,.
\end{equation}
In the above expression, $\delta$ is the declination of the source, $\lambda$ is the geographical latitude of the experiment, and $\alpha_m$ is the maximum hour angle within which a given declination $\delta$ remains inside the field of view of the detector~\cite{Sommers:2000us}. In Tab.~\ref{tab:experiments_parameters} we report the key parameters of all the experiments considered in this work, while in Fig.~\ref{fig:efficiency_curves} we show their geometrical acceptance efficiency $\omega(\delta)$ as a function of the declination. The shape and width of each curve are set by the detector latitude $\lambda$ and zenith cut $\theta_m$, the latter being the maximum zenith angle within which the detector operates at full efficiency. Together, these parameters define the range of declinations that contribute to the exposure.
Including $\omega_{\rm exp}(\delta)$ in the definition of the $\mathcal{D}$-factor is crucial for obtaining reliable bounds on the DM lifetime from UHE gamma-ray surveys. 
Indeed, the response of a ground-based observatory is inherently non-uniform: only directions within its field of view contribute to the detected events at any given time, and the experiments analyzed here exhibit significantly different sky coverages and sensitivities due to their geographical latitudes and zenith-angle acceptances.
Accounting for this directional dependence is essential because the DM-induced gamma-ray flux follows the spatial distribution of the galactic halo and is therefore intrinsically anisotropic~\cite{Dubovsky:1998pu}.}

\textcolor{black}{A direct consequence of this directional dependence is that the SL+IR contribution to the optical depth is negligible for all experiments considered here.
The experiments pointing toward the galactic center, such as PAO and TA, are sensitive only to the high-energy end of the spectrum ($E_\gamma \gtrsim 10^7$~GeV), where the SL+IR term is subdominant compared to the CMB contribution, while those probing lower energies, such as CASA-MIA and KASCADE, do not include the galactic center in their FOV and are thus only marginally affected by the high-energy tail of the SL+IR component. Consequently, we can approximate $\tau_{\gamma\gamma}(E_\gamma, s, b, l) \simeq \tau_{\gamma\gamma}^{\rm CMB}(E_\gamma, s)$, neglecting the SL+IR contribution to the absorption.}
\begin{figure}[t!]
    \centering
    \includegraphics[width=0.48\textwidth]
    {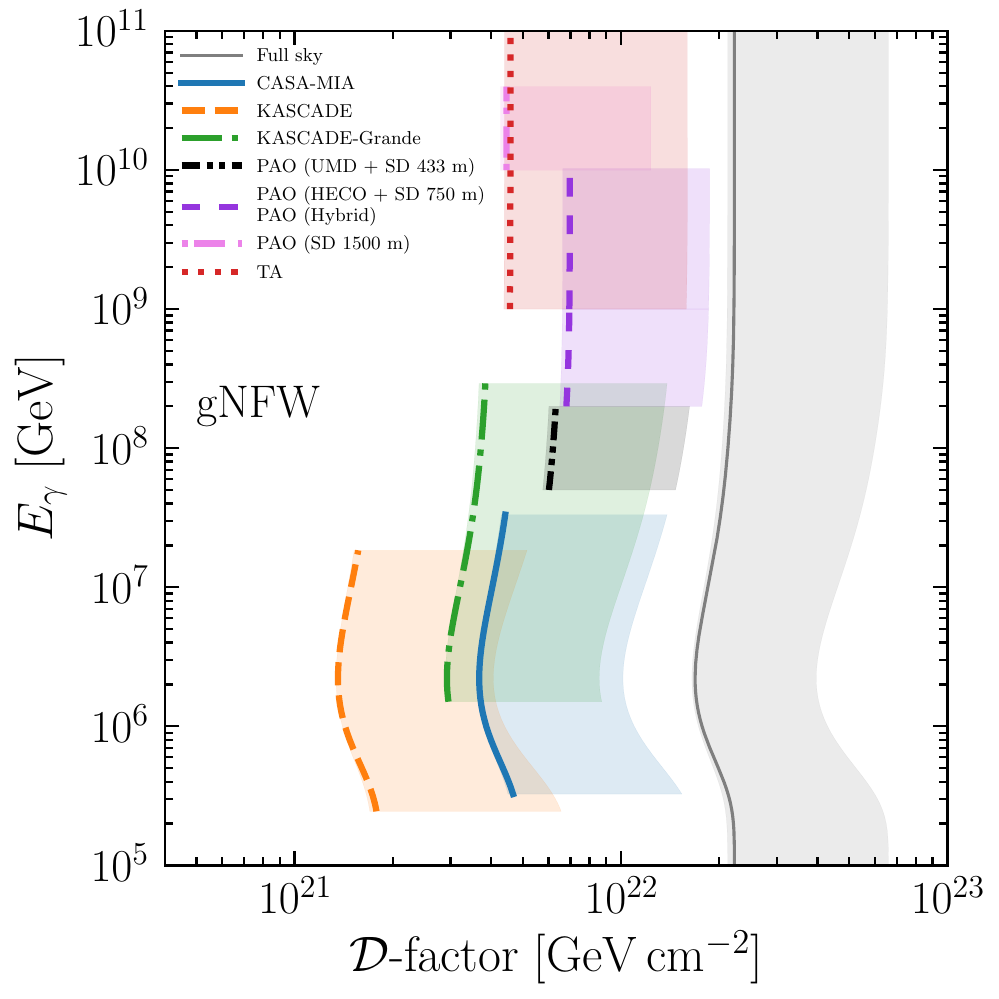}
    \includegraphics[width=0.48\textwidth]
    {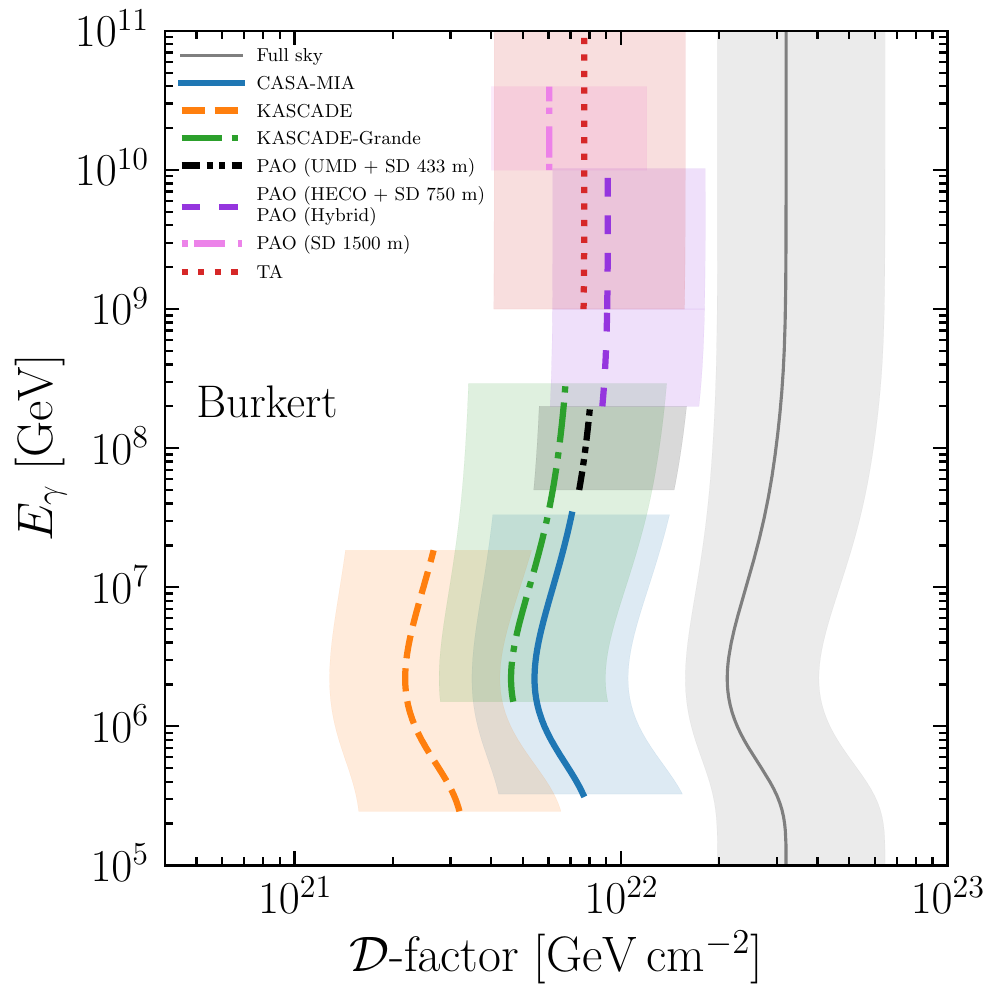}
    \caption{$\mathcal{D}$-factor as a function of the gamma-ray energy computed for the full sky and for the experiments listed in Tab.~\ref{tab:experiments_parameters}, shown with different colors. The x-axis and y-axis are swapped for better visualization. The left and right plots refer to the gNFW and Burkert DM profiles, respectively. The curves with distinct styles correspond to the fiducial DM profiles, while the shaded regions quantify the $2\sigma$ uncertainty on the DM-halo parameters from Ref.~\cite{Benito:2020lgu} (see text and Tab.~\ref{tab:D_factor}).}
    \label{fig:Dfactor}
\end{figure}

\textcolor{black}{In this work, We use the generalized Navarro--Frenk--White (gNFW) and the Burkert profiles as representatives of the two main classes of DM halo models, namely cuspy profiles, which diverge as the galactocentric distance $r$ approaches zero, and cored profiles, which exhibit a finite central density in the same limit, respectively.}
The gNFW profile takes the expression
\begin{equation}\label{eq:gNFW_profile}
        \rho_{\mathrm{gNFW}}(r)= \frac{\rho_{s}}{(r/R_{s})^{\gamma}(1+r/R_{s})^{3-\gamma}} \,,
\end{equation}
where $R_{s}$ is the scale radius, $\rho_{s}$ is a normalization factor, and $\gamma$ is the inner slope of the profile. The inner slope is typically considered to vary from 0.1 to 1.4. Setting $\gamma = 1$, we obtain the commonly-used NFW profile~\cite{Navarro:1996gj}. The Burkert profile is instead defined as~\cite{Burkert:1995yz}
\begin{equation}\label{eq: Burkert_profile}
	\rho_{\text{Burk}}(r)=\frac{\rho_{c}R_{c}^3}{(r+R_{c})(r^2+R_{c}^2)}\,,
\end{equation}
where $R_{c}$ is the core radius and $\rho_c$ denotes the central density. In our study, we employ fiducial values for the parameters of the two profiles, specifically a local DM density $\rho_0 = 0.4~{\rm GeV/cm^3}$ and a characteristic radius $R_s = R_c = 25~{\rm kpc}$. \textcolor{black}{We note that the fiducial local DM density is selected to enable direct comparison with other analyses adopting the same value.}
We assign a fiducial value of $\gamma = 1.0$ to the slope of the gNFW profile, corresponding to the standard NFW profile. The normalization of the DM profiles is then determined from the local density through the relation $\rho_{\rm DM}(R_\odot) = \rho_0$. These parameter selections enable effective comparisons with other studies, as these values are prevalent in the literature. 

In Fig.~\ref{fig:Dfactor} we display the $\mathcal{D}$-factor as a function of the gamma-ray energy for the gNFW and Burkert DM profiles in the left and right panels, respectively. The effect of the gamma-ray attenuation is visible for energies from $\sim 10^5$ to $\sim 10^7$~GeV, where the $\mathcal{D}$-factor deviates from its constant value. \textcolor{black}{In the plot, we compare the $\mathcal{D}$-factors for the different experiments listed in Tab.~\ref{tab:experiments_parameters} with the one computed over the full sky.} The curves correspond to our fiducial choices for the parameters of the two profiles, while the shaded regions highlight the current uncertainty at $2\sigma $ CL on the $\mathcal{D}$-factor from the current measurements of the galactic DM distribution.\footnote{\textcolor{black}{We note that the $2\sigma$ confidence regions in Ref.~\cite{Benito:2020lgu} are truncated and leave $R_s$ and $R_c$ unconstrained from above. To quantify the impact of larger values of $R_s$ and $R_c$, we have recomputed the $\mathcal{D}$-factor adopting values of $R_s$ and $R_c$ corresponding to twice the maximum values reported in Ref.~\cite{Benito:2020lgu}, namely $R_s = 80$~kpc and $R_c = 60$~kpc, respectively. Allowing the local DM density to vary over its full range from minimum to maximum, we find that the $\mathcal{D}$-factor increases by $16\%$ (gNFW) and $95\%$ (Burkert) with respect to the highest values delimiting our uncertainty bands. This would translate into correspondingly stronger constraints on the DM parameter space. Thus, the limits we report should be regarded as conservative in this regard.}}
Indeed, several sources of uncertainty must be taken into account when attempting to constrain the parameters of the DM density profiles with the galactic rotation curves. Our position within the Milky Way leads to uncertainties in the determination of $R_\odot$ and the Sun's circular velocity. Additional uncertainties arise from the limited knowledge of the baryonic mass components (bulge, stellar disk, and gas)~\cite{Pato:2015dua}, the corresponding normalization factors given by the microlensing optical depth for the bulge~\cite{MACHO:2004slp} and the local surface density for the stellar disk~\cite{Bovy:2013raa}, and the systematic errors in the observational determination of the Milky Way rotation curves~\cite{Sofue:2008wu, Sofue:2008wt, Sofue:2013kja}. Hence, in order to take into account the astrophysical uncertainties, we determine the parameters defining the profiles in Eqs.~\eqref{eq:gNFW_profile} and~\eqref{eq: Burkert_profile} according to the $2\sigma$ contours provided in Ref.~\cite{Benito:2020lgu}. Specifically, we use the contours in which the Sun’s galactocentric distance is fixed to the value \textcolor{black}{$R_\odot = 8.178~{\rm kpc}$}, while the remaining parameters are profiled over following a frequentist method (for details, see Refs.~\cite{Rolke:2004mj, Benito:2019ngh}).
\begin{table}[t!]
    \centering
    \hspace*{-0.5em} 
    \begin{tabular}{l @{\hspace{1.5em}} c @{\hspace{1.5em}} c @{\hspace{1.5em}} c @{\hspace{1.5em}} c} 
        \toprule 
         & \textbf{Burkert profile} &  & \textbf{gNFW profile}& \\
           Full-sky $\mathcal{D}$-factor & \multirow{2}{4em}{} & \multirow{2}{4em}{$~\gamma = 0.1$} & \multirow{2}{4em}{$~\gamma=1.0$} & \multirow{2}{4em}{$~\gamma=1.4$}\\
           $[10^{23}~{\rm GeV\, cm^{-2}}]$ &&&&\\
        \midrule 
        Min value & $2.48$ & $2.67$ & $2.76$ & $2.86$\\
        Max value & $8.10$ & $8.29$& $5.71$ & $4.46$\\
        Fiducial value & $4.03$ & $-$ & $2.80$ & $-$\\
        \midrule
    \end{tabular}
    \hspace*{-0.5em} 
    \caption{Minimum, maximum and fiducial values of the full-sky $\mathcal{D}$-factor at $E_\gamma = 10^{9}~{\rm GeV}$ for the Burkert and gNFW DM profiles according to the $2\sigma$ uncertainty from the determination of the galactic DM distribution~\cite{Benito:2020lgu}. The fiducial values correspond to a local DM density of $\rho_0 = 0.4~{\rm GeV/cm^3}$ and a characteristic radius $R_s = R_c = 25~{\rm kpc}$.}
\label{tab:D_factor}
\end{table}

We find that taking into account the FOV and the geometrical acceptance efficiency of each experiment reduces the $\mathcal{D}$-factor up to an order of magnitude (see the dashed orange curve for the KASCADE experiment). As will be discussed later, this result has a strong impact on the constraints on the DM parameter space. On the other hand, the uncertainty in the DM profile leads only to variations of a few in the $\mathcal{D}$-factor. This is also evident in Tab.~\ref{tab:D_factor} where we report the minimum, maximum and fiducial values of the full-sky $\mathcal{D}$-factor obtained by varying the DM-halo parameters within their $2\sigma$ uncertainties and fixing the gamma-ray energy at $E_\gamma = 10^9~{\rm GeV}$ (no attenuation). We emphasize that for the gNFW profile the minimum and the maximum values are both attained at a slope value of $\gamma = 0.1$, for which large values for the scale radius $R_s$ are allowed. Moreover, the fiducial NFW profile yields a $\mathcal{D}$-factor that is nearly identical to the minimum gNFW value, thus explaining the curves in the left panel of Fig.~\ref{fig:Dfactor}.

%========================================
\section{Constraints on the dark matter lifetime}
\label{sec:constraints}
%========================================
\begin{figure}[t!]
    \centering
    \includegraphics[width=0.46\linewidth]{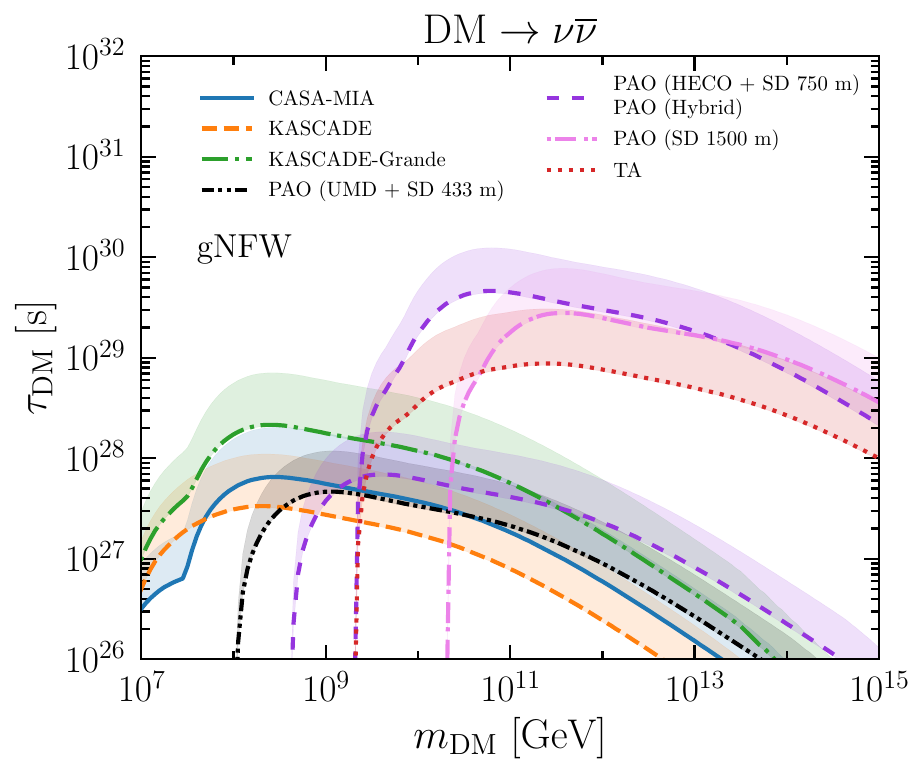}
    \hspace{0.04\linewidth}
    \includegraphics[width=0.46\linewidth]{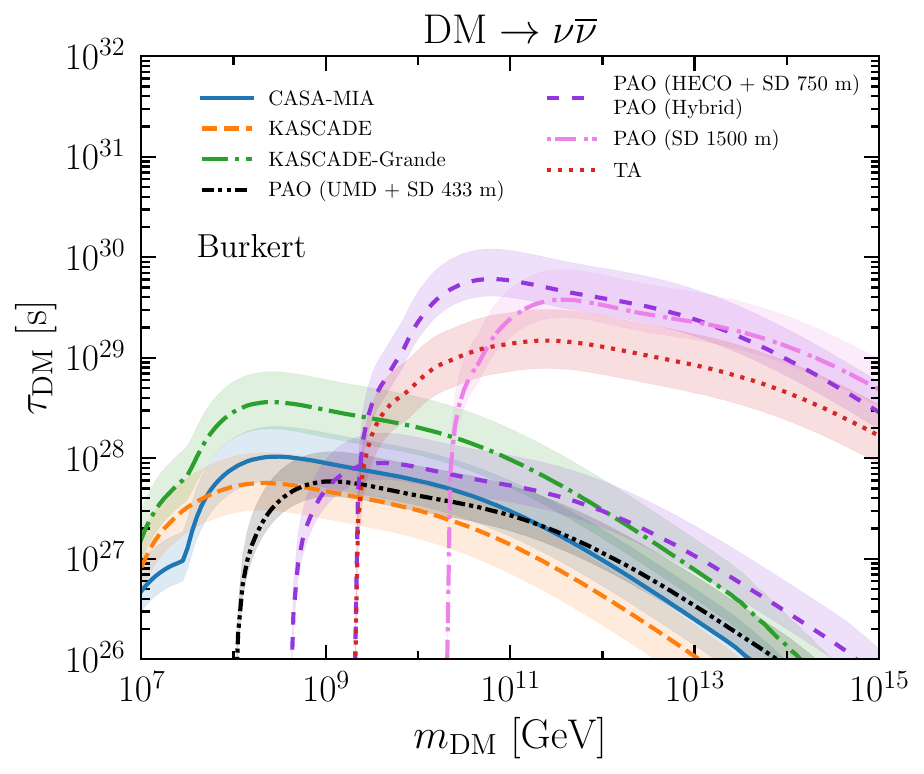}
    \\[0.05cm]
    \includegraphics[width=0.46\linewidth]{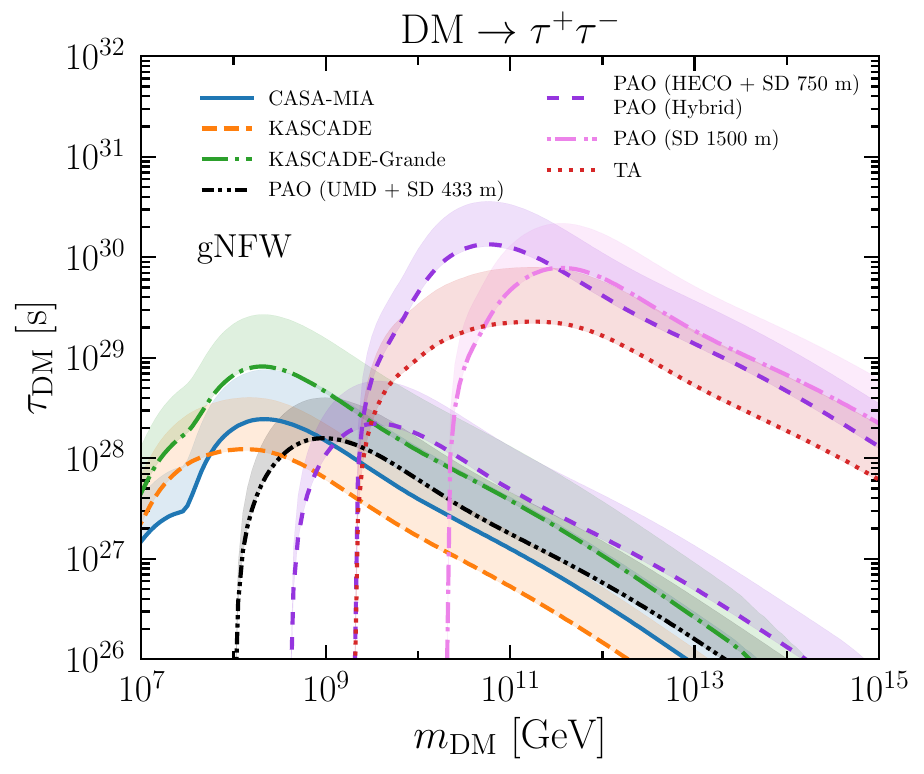}
    \hspace{0.04\linewidth}
    \includegraphics[width=0.46\linewidth]{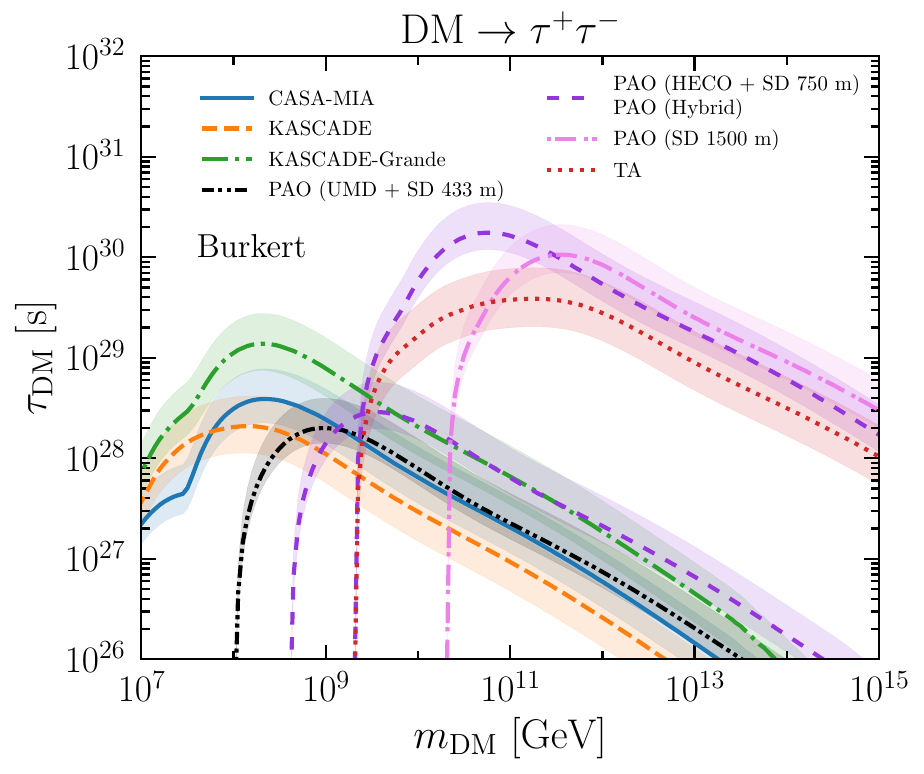}
    \\[0.05cm]
    \includegraphics[width=0.46\linewidth]{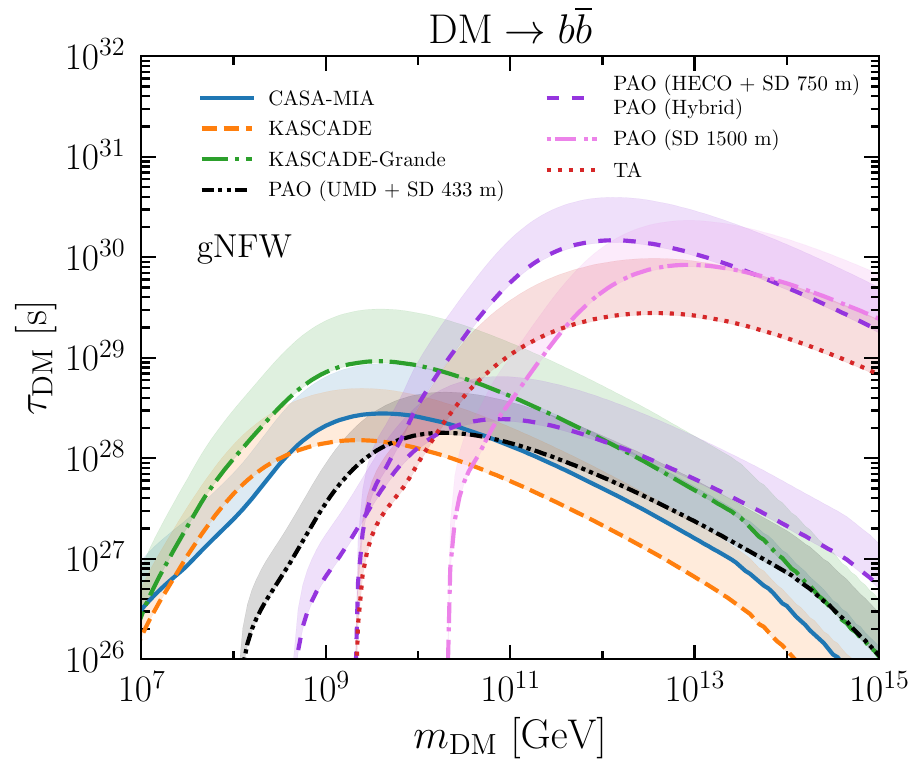}
    \hspace{0.04\linewidth}
    \includegraphics[width=0.46\linewidth]{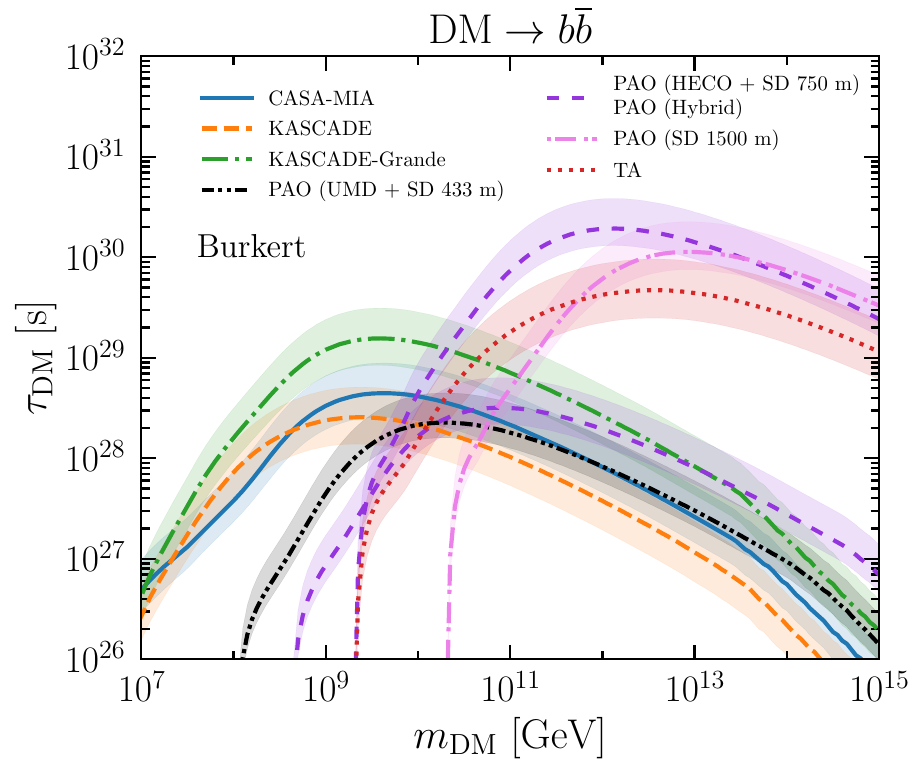}
    \caption{Lower bounds on the lifetime of decaying DM particles at 95\% CL for the channels ${\rm DM} \to \nu \overline{\nu}$ (upper plots), ${\rm DM} \to \tau^+ \tau^-$ (middle plots) and ${\rm DM} \to b \overline{b}$ (lower plots) assuming the gNFW (left plots) and Burkert (right plots) profiles. Different styles correspond to the datasets reported in Tab.~\ref{tab:experiments_parameters}. The lines show the constraints assuming the fiducial values for the DM profiles, while the bands cover the $2\sigma$ uncertainty on the $\mathcal{D}$-factor (see Fig.~\ref{fig:Dfactor}).}
    \label{fig:constraints}
\end{figure}

We derive the limits on the parameter space of heavy decaying DM particles by comparing the predicted integral $\gamma$-ray flux with the observational upper bounds. Following the same procedure adopted in Ref.~\cite{Chianese:2021jke}, for each dataset reported in Tab.~\ref{tab:experiments_parameters}, we consider the test statistic (TS):
\begin{equation}\label{eq:chi2}
    \mathrm{TS}(m_{\mathrm{DM}},\tau_{\mathrm{DM}}) = \sum_{i=1}^{N}\left(\frac{\Phi_{\gamma}(E_{\gamma, i} | m_{\mathrm{DM}},\tau_{\mathrm{DM}}) -\Phi^{\mathrm{data}}_{\gamma,i}}{\sigma_i}\right)^{2}\,,
\end{equation}
where $\Phi_{\gamma}(E_{\gamma, i} | m_{\mathrm{DM}},\tau_{\mathrm{DM}})$ is the DM–induced integral flux of Eq.~\eqref{eq:integral}, while $\Phi^{\mathrm{data}}_{\gamma,i}$ and $\sigma_i$ represent, respectively, the reported flux values and their uncertainties. Each term in the sum corresponds to one experimental energy bin. Because the measurements are published as upper limits, we simply take $\Phi^{\mathrm{data}}_{\gamma,i}=0$, and the quantities $\sigma_i$ are reconstructed at the $68\%$~CL according to the confidence level associated with each experimental constraint. 

We set the $95\%$~CL lower bound on the lifetime for DM masses from $10^7$ to $10^{15}$~GeV by assuming that any potential signal originates solely from DM decay. The requirement that the hypothesis of zero observed photon events is excluded with more than $95\%$ probability translates into the condition $\mathrm{TS}=2.76$. This threshold does not come from the standard $\chi^2$ distribution because the flux measurements are restricted to be non-negative. Accounting for this constraint leads to the modified probability density
\begin{equation}
    P(x)=\frac{e^{-x/2}}{\sqrt{2\pi x}} \frac{1+\Theta(\Lambda-x)} {1+\mathrm{Erf}(\Lambda/\sqrt{2})}\,,
\end{equation}
with $x=\mathrm{TS}$ and $\Lambda=\sum_{i=1}^{N}
\left[{\Phi_{\gamma}  (E_{\gamma, i} | m_{\mathrm{DM}},\tau_{\mathrm{DM}})}/{\sigma_i}\right]^2$, which corresponds to the TS evaluated at $\Phi^{\mathrm{data}}_{\gamma,i}=0$. The Heaviside function $\Theta$ appears because of the positivity constraint on the measured fluxes. In the limit of large $\Lambda$, when the predicted fluxes are well above zero, the expression converges to the usual $\chi^2$ behavior. Since in our case the measured TS is equal to $\Lambda$, the $95\%$ exclusion value is obtained by imposing $\Lambda=2.76$.

In Fig.~\ref{fig:constraints}, we show the 95\% confidence-level lower limits on the lifetime of decaying DM particles for three representative decay channels, organized into upper, middle, and lower panels corresponding to ${\rm DM} \to \nu \overline{\nu}$, ${\rm DM} \to \tau^+ \tau^-$, and ${\rm DM} \to b \overline{b}$, respectively (see Appendix~\ref{app:channels} for other decay channels). Each channel is evaluated under two benchmark assumptions for the galactic DM distribution: a cuspy gNFW profile (left column) and a cored Burkert profile (right column). The various line styles reflect the different observational datasets summarized in Tab.~\ref{tab:experiments_parameters}, enabling a clear comparison of their relative constraining power over the full DM mass range. The different curves represent the limits obtained using the fiducial values of the respective halo profiles, while the surrounding shaded bands illustrate the astrophysical uncertainties by spanning the $2\sigma$ range of the corresponding $\mathcal{D}$-factor, as derived in the previous Section. This figure highlights both the dependence of the inferred limits on the assumed galactic density profile and the degree to which halo-model uncertainties propagate into the final constraints.
\begin{figure}[t!]
    \centering
    \includegraphics[width=0.32\linewidth]{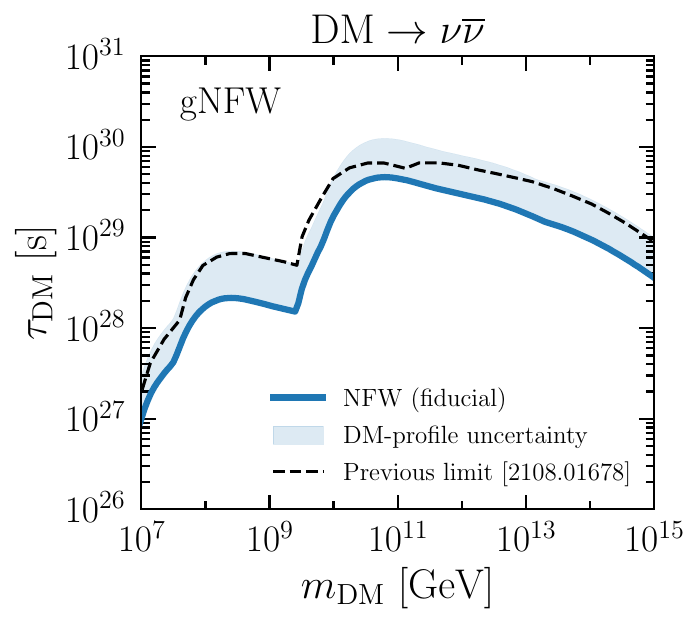}
    \includegraphics[width=0.32\linewidth]{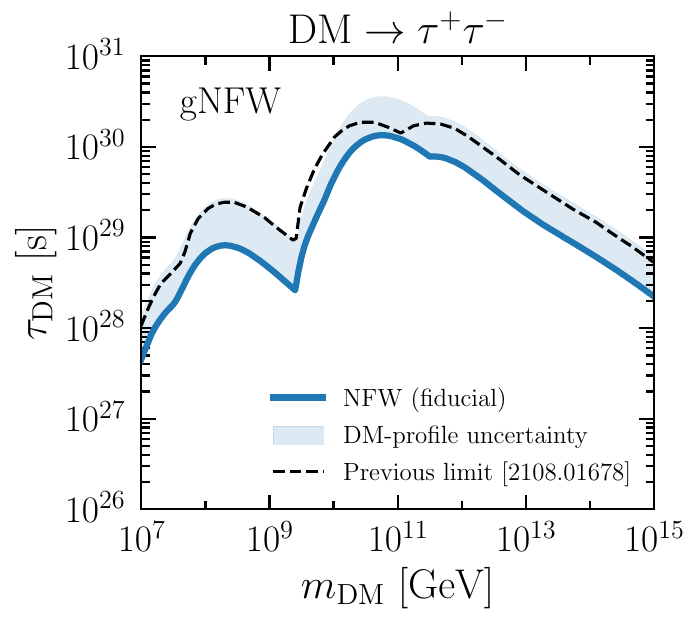}
    \includegraphics[width=0.32\linewidth]{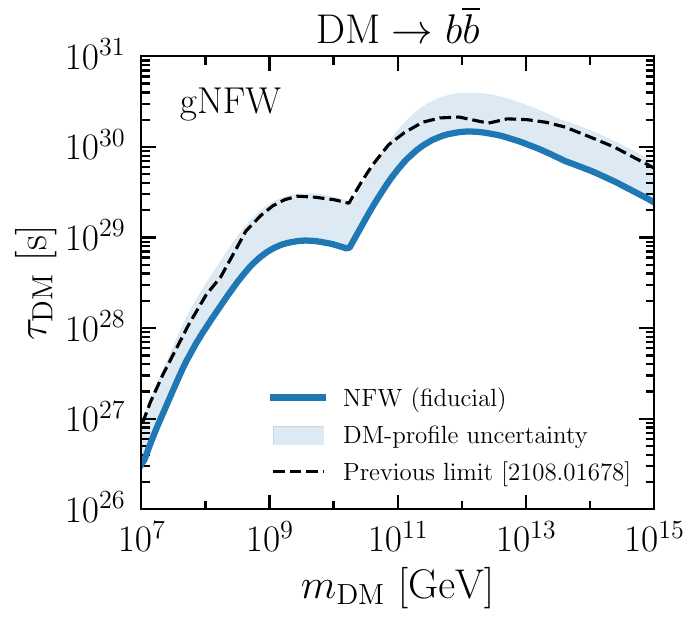}
    \caption{Comparison between the overall UHE gamma-ray bounds obtained in the present work (blue solid lines) and the ones obtained in Ref.~\cite{Chianese:2021jke} (black dashed lines), considering different DM decay channels and assuming the fiducial NFW profile (i.e. gNFW with slope $\gamma=1$, see Eq.~\eqref{eq:gNFW_profile}) with $\rho_0 = 0.4~{\rm GeV/cm^3}$ and $R_s = 25.0~{\rm kpc}$. The shaded regions correspond to the $2\sigma$ uncertainty according to the current determination of the parameters of the gNFW profile.}
    \label{fig:comparison}
\end{figure}
In Fig.~\ref{fig:comparison}, we show a direct comparison between the overall UHE gamma-ray constraints derived in the present work and the corresponding limits reported in Ref.~\cite{Chianese:2021jke}. The blue solid lines represent our updated results, while the black dashed lines indicate the previous bounds. The figure considers the three representative decay channels, all evaluated assuming the fiducial NFW profile. As for the previous figure, the shaded regions surrounding the curves denote the $2\sigma$ uncertainty on the limits, which arises from the current determination of the parameters of the gNFW profile. This comparison clearly illustrates the improvements achieved with the present analysis. Specifically, we find that, despite considering a higher DM halo density and the updated limits on the UHE gamma-ray diffuse flux (e.g. from TA and PAO), the resulting bounds are weaker due to the inclusion of the geometrical acceptance efficiency of each experiment. The figure also shows how the derived limits depend on the choice of halo-profile parameters, indicating that uncertainties in the DM distribution affect the robustness of the constraints.
\begin{figure}[t!]
    \centering
    \includegraphics[width=0.80\linewidth]{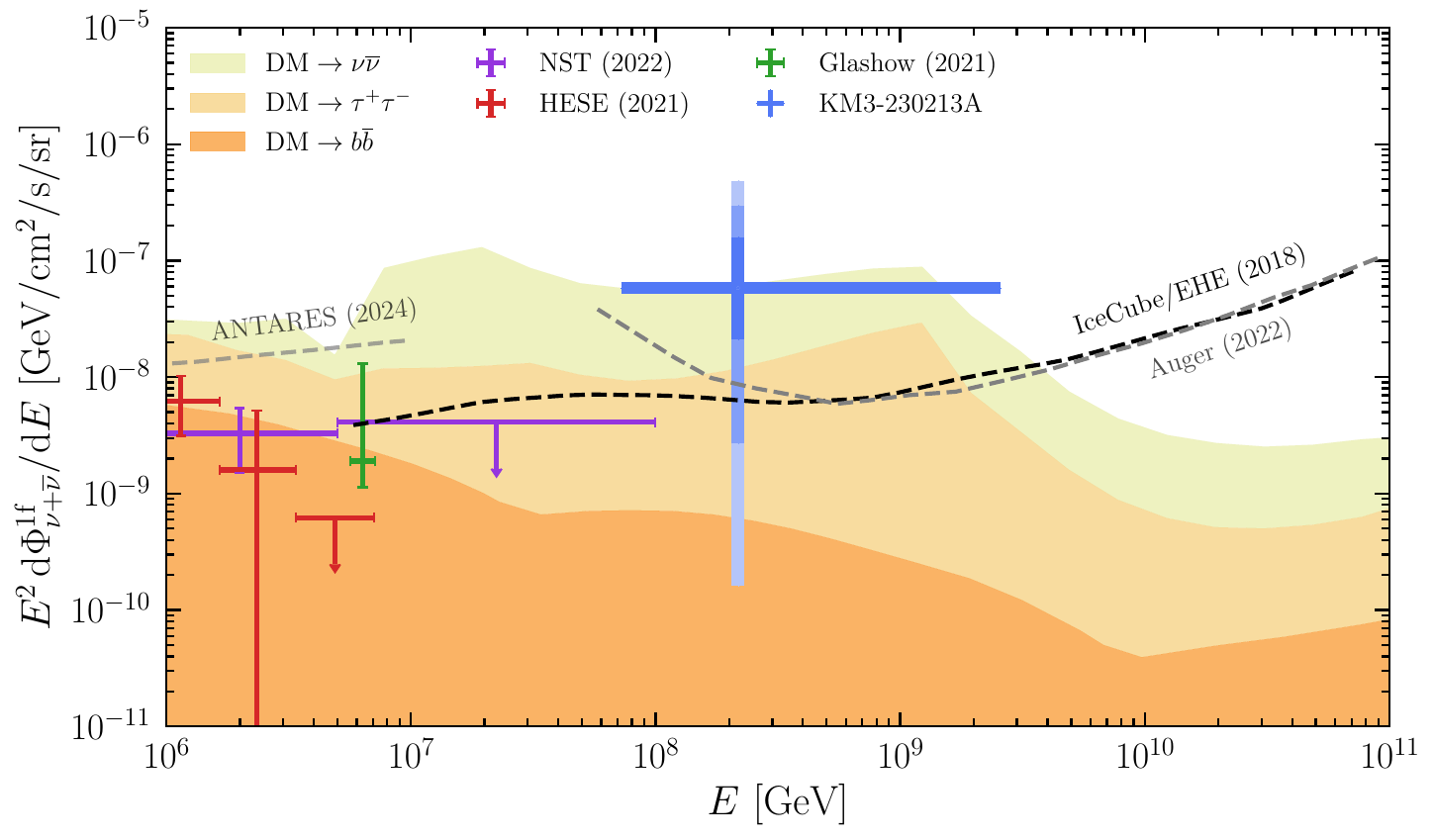}
    \caption{Maximum all-sky single-flavor neutrino and antineutrino flux from the decays of SHDM particles as constrained by gamma-ray data. The shaded regions represent the overlap of the neutrino fluxes related to different SHDM masses ranging from $10^6$ to $10^{15}$~GeV for different decay channels. Also shown are the IceCube data below $10^8~{\rm GeV}$~\cite{Abbasi:2021qfz, IceCube:2020wum, IceCube:2021rpz}, the flux estimate from the KM3-230213A event~\cite{KM3NeT:2025npi}, and the upper bounds (dashed lines) placed by ANTARES~\cite{ANTARES:2024ihw}, IceCube/EHE~\cite{IceCube:2018fhm} and Pierre Auger~\cite{PierreAuger:2023pjg} (from left to right).}
    \label{fig:neutrinos}
\end{figure}

In Fig.~\ref{fig:neutrinos}, we show the maximum all-sky single-flavor neutrino and antineutrino flux that is allowed by the gamma-ray limits on the SHDM particle decays. The neutrino flux is computed by summing the galactic and extragalactic components and averaging over the full sky. The shaded regions indicate the envelope resulting from the superposition of the neutrino fluxes corresponding to SHDM masses in the range $10^6$-$10^{15}$~GeV. For each SHDM mass, we consider the lowest allowed values for the SHDM lifetime according to the bounds derived in the present work and to the ones from LHAASO Collaboration~\cite{LHAASO:2022yxw} after a proper rescaling of the DM halo density to the current estimate. These regions therefore illustrate the theoretical uncertainty associated with the SHDM parameter space and the decay-mode channels. For comparison, we also include existing experimental measurements and limits: the IceCube neutrino flux measurements (the Northern Sky Tracks (NST), the High-Energy Starting Events (HESE), and the Glashow resonance event) at energies below $10^8~\,\mathrm{GeV}$~\cite{Abbasi:2021qfz, IceCube:2020wum, IceCube:2021rpz}, the flux estimate inferred from the KM3-230213A event~\cite{KM3NeT:2025npi}, and the upper bounds (shown as dashed lines) obtained by ANTARES~\cite{ANTARES:2024ihw}, IceCube in the Extremely High-Energy (EHE) regime~\cite{IceCube:2018fhm}, and the Pierre Auger Observatory~\cite{PierreAuger:2023pjg}. The comparison highlights the interplay between gamma-ray and neutrino observations in probing the SHDM decay scenarios. Specifically, hadronic decay channels (e.g. ${\rm DM} \to b\overline{b}$) result to be more constrained by the gamma-ray data while being consistent with current neutrino data. On the other hand, the leptonic channels (e.g. ${\rm DM} \to \tau^+\tau^-$ and ${\rm DM} \to \nu\overline{\nu}$) are better probed by neutrino measurements especially for SHDM masses below $10^8~{\rm GeV}$. Interestingly, we find that the leptonic channels are compatible with the flux expected from the KM3-230213A event (a dedicated analysis is beyond the scope of the present work).

%========================================
\section{Conclusions}
\label{sec:conclusions}
%========================================

The indirect approach based on the observation of high-energy gamma rays produced by the decay of dark matter particles constitutes an important strategy for placing stringent constraints on the dark matter parameters. In this regard, we have re‑examined the constraints on super‑heavy dark matter lifetimes, in the mass range from $10^7$ to $10^{15}$~GeV, using the ultra-high-energy gamma‑ray searches carried out by the CASA-MIA, KASCADE, KASCADE-Grande, Pierre Auger Observatory and Telescope Array experiments, placing emphasis on realistic instrument response and astrophysical uncertainties. In particular, by folding the field‑of‑view and the geometrical acceptance of each experiment into the analysis, we obtain dark matter lifetime limits that are slightly more conservative than those in previous works but substantially more faithful to the true sensitivity of current air‑shower arrays. Indeed, accounting for the directional dependency is crucial when analyzing the anisotropic gamma-ray fluxes expected from dark matter decays, which are influenced by the spatial distribution of the galactic halo. Across a variety of two‑body decay channels and halo models (NFW and Burkert profiles), the present ultra-high-energy gamma‑ray data exclude sizeable regions of the parameter space of super-heavy dark matter particles and complement limits from neutrinos. Importantly, our results elucidate the potential tension between the upper bounds on the gamma-ray emission and the possible interpretation of the KM3-230213A event in the context of decaying super-heavy dark matter models.

%========================================
\section*{Acknowledgements}
%========================================

We acknowledge the support by the following research projects funded by the Istituto Nazionale di Fisica Nucleare (INFN): MC and NS by the TAsP (Theoretical Astroparticle Physics) project; SC and FS by the QGSKY (Quantum Gravity in the SKY) project; VN by the MoonLIGHT-2 project; and AT by the TEONGRAV project. The work of NS is further supported by the research grant number 2022E2J4RK ``PANTHEON: Perspectives in Astroparticle and Neutrino THEory with Old and New messengers'' under the program PRIN 2022 funded by the Italian Ministero dell’Università e della Ricerca (MUR).

This work has been carried out in the context of the course ``Introduction to Astroparticle Physics'', taught by MC and NS, and offered by the PhD program ``Cosmology, Space Science \& Space Technology (SPACE)'' at the Scuola Superiore Meridionale (SSM). The numerical analysis was performed by the PhD students SC, VMG, VN, FS, and AT, who were organized into independent groups, each dedicated to a specific section. The entire project was coordinated under the supervision of MC and NS, who collected and checked all the numerical codes developed by the students. All the authors participated in writing the manuscript and preparing the figures.

\bibliography{bibliography}

\newpage
\appendix

%========================================
\section{Geometrical acceptance efficiency of the experiments}
\label{sec:efficiency}
%========================================

\textcolor{black}{In this appendix we report the full derivation of the geometrical acceptance efficiency given in Eq.~\eqref{eq:omega_delta}. Specifically, we extend the treatment of  Ref.~\cite{Sommers:2000us} and compute the normalization factor of $\omega_{\rm exp}(\delta)$, which is essential in our treatment, where the acceptance efficiency enters as a weighting factor in the very definition of the 
$\mathcal{D}$-factor given in Eq.~\eqref{eq:Dfactor}.
For a detector located at latitude $\lambda$, we adopt the standard  assumptions employed in wide-field, ground-based EAS experiments~\cite{Sommers:2000us, Esmaili:2015xpa}:
\begin{enumerate}
    \item \emph{Full-time operation}: the detector acquires data over a full sidereal day;
    \item \emph{Azimuthal symmetry}: the aperture is uniform with respect to azimuth (or right ascension in equatorial coordinates);
    \item \emph{Separable geometric term}: although the instantaneous aperture generally depends on both the arrival direction and the event energy, we adopt a simplified treatment in which the geometrical contribution is factorized from the detector energy response.
    \item \emph{Full efficiency} within a zenith cut $\theta_m$ characteristic of the experiment.
\end{enumerate}
Under these considerations, the sole factor governing the instantaneous aperture $A_{\text{eff}}$ is the geometric projection $\cos\theta_z(t)$, thus having 
\begin{equation}
A_{\text{eff}}(\theta_z(t)) = A_0 \cos\theta_z(t)\,_{\{\theta_z(t)\le\theta_m\}}\,,
\label{eq:Aeff_simplified}
\end{equation}
where $A_0$ is an effective constant
geometrical area and the condition $\theta_z(t)<\theta_m$ encodes the visibility of the source within the FOV of the detector.} From the instantaneous aperture, we can compute the exposure $\mathcal{E}(\alpha,\delta)$ for a source at right ascension $\alpha$ and declination $\delta$ as~\cite{Sommers:2000us}
\begin{equation}
\mathcal{E}(\alpha,\delta) = \int A_{\text{eff}}(t)\,{\rm d}t\,.
\end{equation}
This function quantifies both the duration and efficacy of the detector’s observation of a specific direction, incorporating the Earth’s rotational dynamics. 
In the case of a continuously operating, azimuthally symmetric array, all sources situated along the same declination circle are observed identically  throughout a sidereal day. This indicates that the exposure is invariant with respect to the parameter  $\alpha$, and it suffices to evaluate only $\mathcal{E}(\delta)$.
By substituting time with the local hour angle $H$, under the transformation ${\rm d}H = \omega_{\mathrm{Earth}}\,{\rm d}t$, and limiting the integral to those intervals when the source remains within the zenith cut ($\theta_z \le \theta_m$), we obtain: 
\begin{equation}
\mathcal{E}(\delta) = \frac{A_0}{\omega_{\mathrm{Earth}}} \int_{H\in\text{visible}} \cos\theta_z(\lambda,\delta;H)\,{\rm d}H\,,
\label{eq:Eraw}
\end{equation}
where $\omega_{\mathrm{Earth}}$ is the Earth’s sidereal angular velocity.
\begin{figure}[t!]
    \centering
    \includegraphics[width=0.48\textwidth]{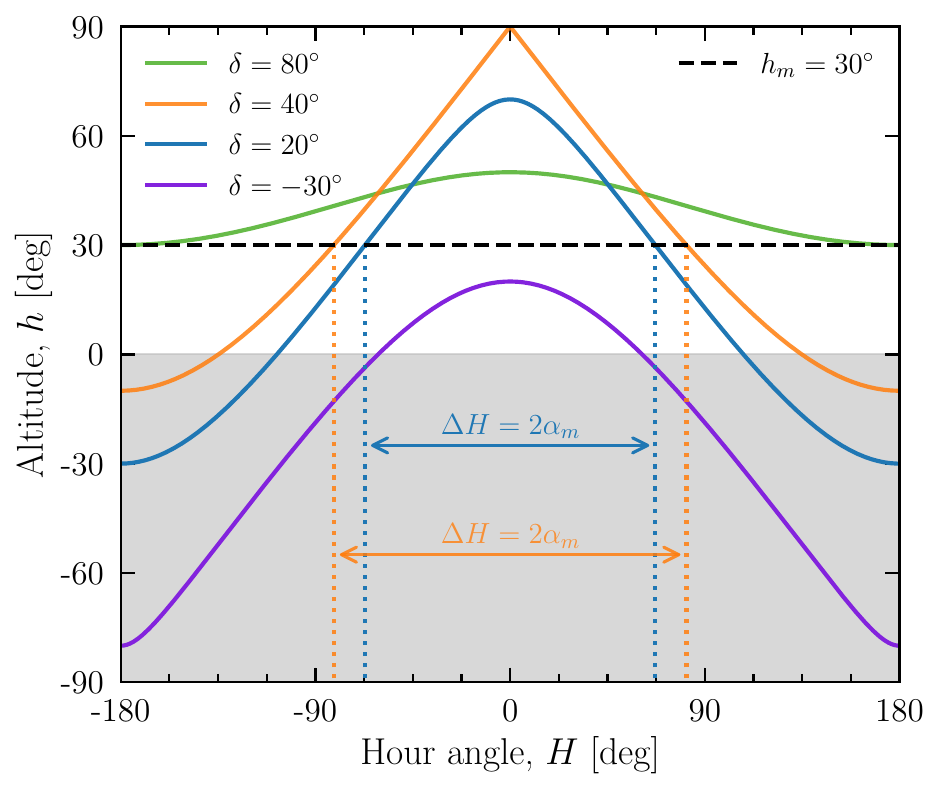}
    \includegraphics[width=0.48\textwidth]{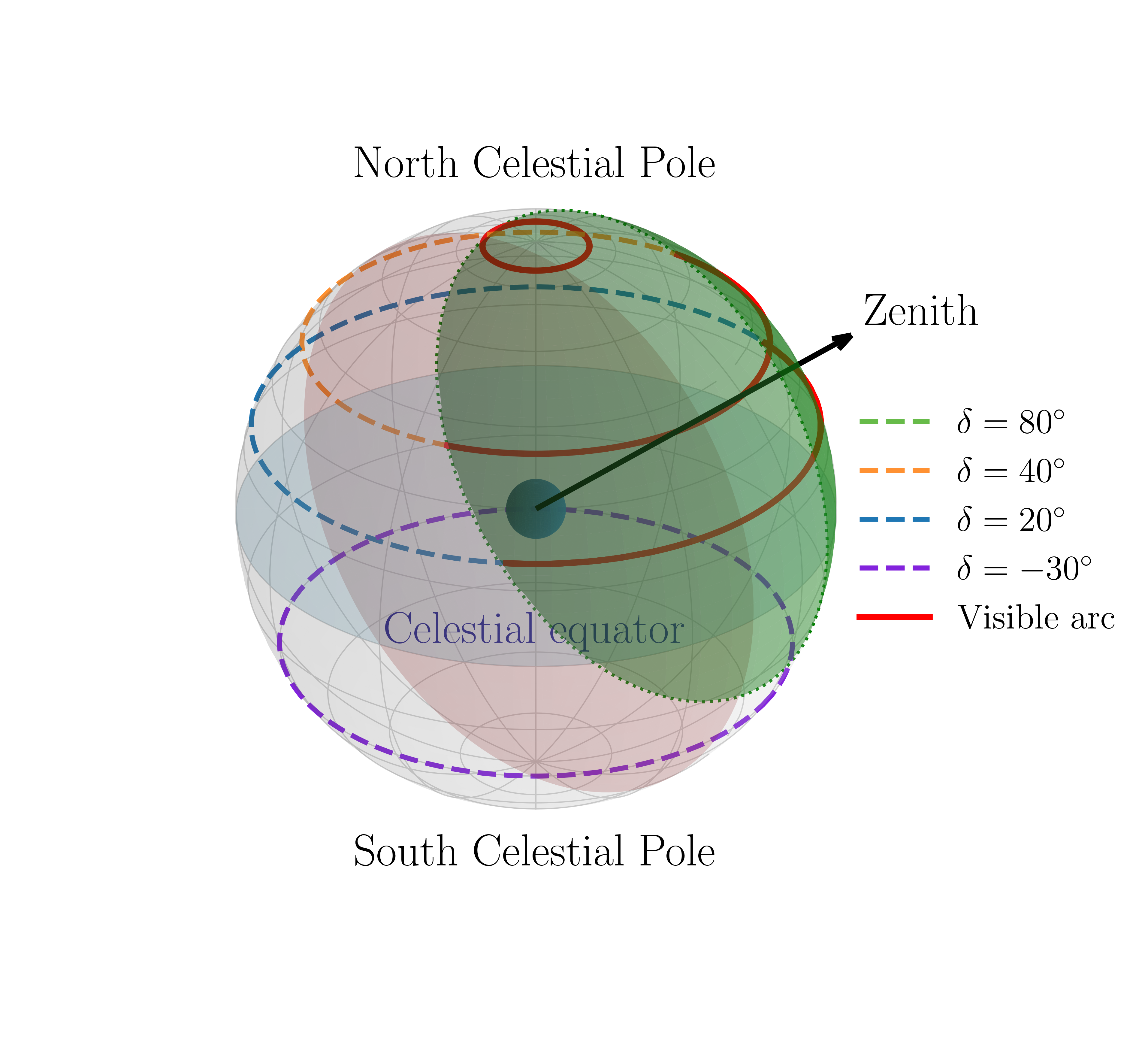}
    \caption{Geometrical interpretation of the visibility condition for a detector located at latitude $\lambda = 40^{\circ}$ and with a zenith cut $\theta_m = 60^{\circ}$. The left plot shows the altitude curves and the corresponding visible hour-angle intervals for selected declinations. The visibility interval $\Delta H = 2\alpha_m$ corresponds to the portion of trajectory lying above the minimum altitude $h_m$. The graphic on the right shows the same trajectories projected on the celestial sphere together with the detector FOV (green shaded area). The red arc denotes the portion of each trajectory lying above the minimum altitude and thus contributing to the exposure.}
    \label{fig:Altitude(H)&3D_FOV}
\end{figure}
From spherical trigonometry, the projection factor $\cos\theta_z$ can be expressed as a function of the hour angle as~\cite{Green:1985}
\begin{equation}
\sin h(\lambda, \delta; H)=\cos\theta_z(\lambda, \delta; H) =  \sin\lambda\,\sin\delta + \cos\lambda\,\cos\delta\,\cos H\,.
\label{eq:costhetaz}
\end{equation}
A source is visible as long as its altitude $h=90^\circ-\theta_z$ exceeds the minimum altitude $h_m=90^\circ-\theta_m$, leading to the following condition
\begin{equation}
\cos H \ge 
\frac{\cos\theta_m - \sin\lambda\,\sin\delta}{\cos\lambda\,\cos\delta}\,.
\end{equation}
We define $\alpha_m$ as the maximum hour angle at which the celestial source remains within the field of view of the experiment positioned at latitude $\lambda$ on Earth:
\begin{equation}
    \alpha_m = \cos^{-1}\left[\frac{\cos\theta_m -\sin\lambda\sin\delta}{\cos\lambda\cos\delta}\right] = \cos^{-1}\xi\,.
    \label{eq:alpha_m}
\end{equation}
 Consequently, the observable range of hour angles is given by  $H \in [-\alpha_m, +\alpha_m]$. Therefore, $\alpha_m$ quantifies the half-arc of visibility with the limits~\cite{Sommers:2000us}:
\begin{equation*}
\begin{cases}
\alpha_m = 0 & \text{if} \quad \xi >1 \quad \text{(the source is never observable)}
\\ \alpha_m = \pi & \text{if} \quad \xi<-1 \quad \text{(the source is circumpolar, i.e. always observable)}
\\ \alpha_m = \cos^{-1}\xi & \text{if} \quad-1<\xi<1.
\end{cases}
\end{equation*}
Moreover, from Eq.~\eqref{eq:costhetaz} we observe that, when the source passes the local meridian corresponding to $H = 0$, it reaches the maximum altitude $h_{\mathrm{max}}$ equal to
\begin{equation}
    h_{\mathrm{max}} = \sin^{-1}\left(\cos(\delta-\lambda)\right)\,.
\end{equation}
Imposing the visibility condition $h_{\mathrm{max}} \ge 90^{\circ} - \theta_m$, the range of declination circles that fall within the FOV of an experiment is 
\begin{equation}\label{eq:FOV_declination}
    \lambda-\theta_m \le \delta \le \lambda + \theta_m \,.
\end{equation}
In particular, declinations satisfying $\delta =\lambda$ correspond to trajectories that reach the maximum elevation accessible from the detector site. Such transits occur at the smallest zenith angles and therefore benefit from the most favorable observing conditions. The geometrical role of $\alpha_m$, the detector FOV and the geometric considerations discussed above are illustrated in Fig.~\ref{fig:Altitude(H)&3D_FOV}. The plot on the left shows how different declinations produce different altitude tracks and visibility intervals, while the graphic on the right illustrates the corresponding geometric interpretation on the celestial sphere.
\begin{figure}[t!]
    \centering
    \includegraphics[width=1.0\linewidth]{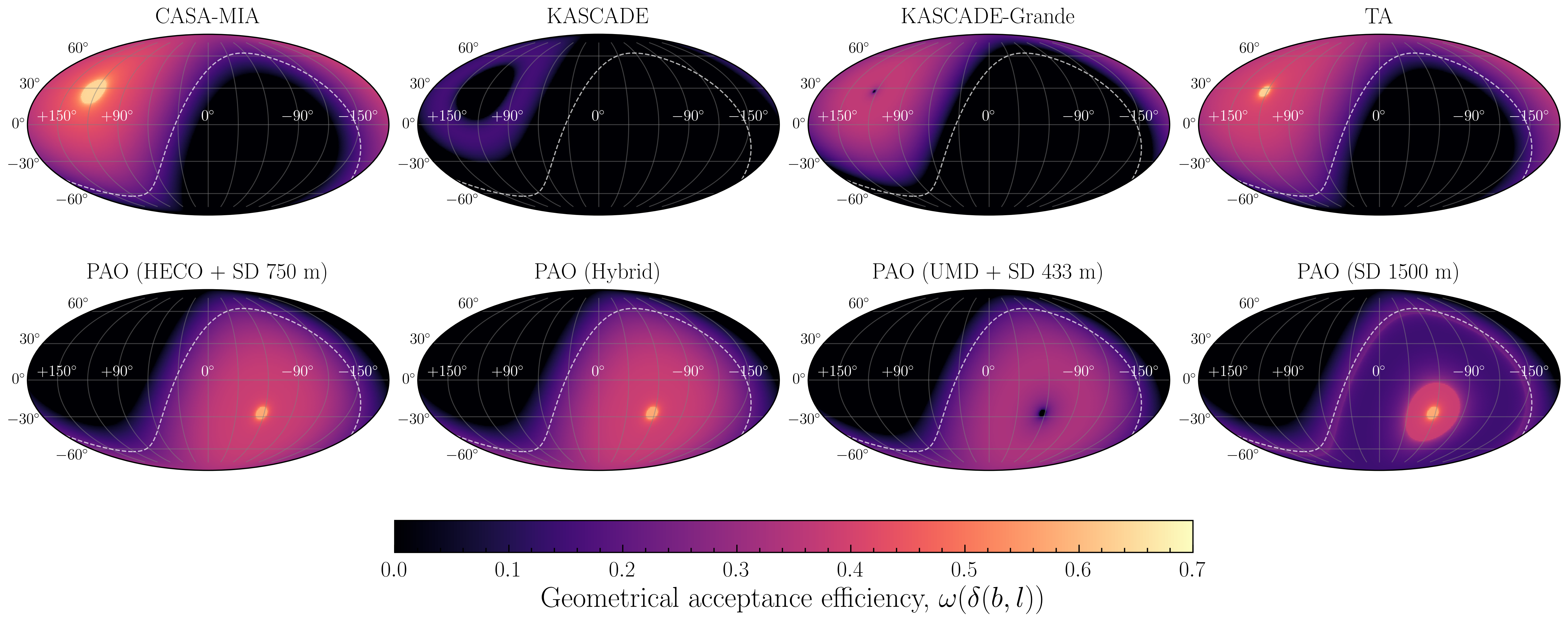}
    \caption{FOV in galactic coordinates for each experiment considered in this study, with color grading indicating the geometrical acceptance efficiency $\omega (\delta(b,l))$ defined in Eq.~\eqref{eq:omega_delta}.}
    \label{fig:FOV_mollweide_weighted}
\end{figure}

Substituting Eq.~\eqref{eq:costhetaz} into Eq.~\eqref{eq:Eraw} and integrating over the visible range yields
\begin{equation}
    \mathcal{E}(\delta) = \frac{A_0}{\omega_{\mathrm{Earth}}}\int_{-\alpha_m}^{+\alpha_m}\cos\theta_z(\lambda, \delta; H) \,{\rm d}H = \frac{2A_0}{\omega_{\mathrm{Earth}}}(\alpha_m\sin\lambda\sin\delta+\cos\lambda\cos\delta\sin\alpha_m).
    \label{eq:E_delta}
\end{equation}
We can now introduce the geometrical acceptance efficiency $\omega_{\rm exp}(\delta)$ as the function that quantifies the relative exposure of a ground-based detector at latitude $\lambda$ to a circle of declination $\delta$~\cite{Sommers:2000us}. We define it by normalizing the exposure to its maximum possible value on the celestial sphere, corresponding to an ideal circumpolar source that remains constantly at the local zenith ($\cos\theta_z = 1$). From Eq.~\eqref{eq:E_delta}, this yields 
\begin{equation}
    \mathcal{E}_{\mathrm{max}} = \frac{A_0}{\omega_{\mathrm{Earth}}}2\pi\,.
\end{equation}
Therefore, the normalized exposure becomes 
\begin{equation}\label{eq:omega_delta_app}
    \omega_{\rm exp}(\delta) = \frac{\mathcal{E}(\delta)}{\mathcal{E}_\mathrm{max}}=\frac{1}{\pi}(\alpha_m \sin\lambda\sin\delta + \cos\lambda \cos\delta \sin\alpha_m)\,, 
\end{equation}
where $\alpha_m$ is given by Eq.~\eqref{eq:alpha_m}. Compared to the expression given in Ref.~\cite{Sommers:2000us}, Eq.~\eqref{eq:omega_delta_app} is by construction $0\le \omega(\delta) \le 1$. Its value reflects both how long a source stays within the FOV and how favorably it transits in altitude. Sources that pass closer to the local zenith contribute more significantly than those remaining near the visibility limit, even for comparable visibility times. This refinement provides an essential geometrical weighting factor for the calculation of fluxes in realistic models of experiments monitoring large portions of the sky. In Fig.~\ref{fig:FOV_mollweide_weighted} we show the quantity $\omega(\delta)$ as a function of the galactic coordinates $b$ and $l$, thus clearly indicating the FOV of each experiment as the colored regions.

\newpage
%========================================
\section{Lifetime bounds for several decay channels \label{app:channels}}
%========================================

\begin{figure}[h!]
    \centering
    \includegraphics[width=0.32\linewidth]{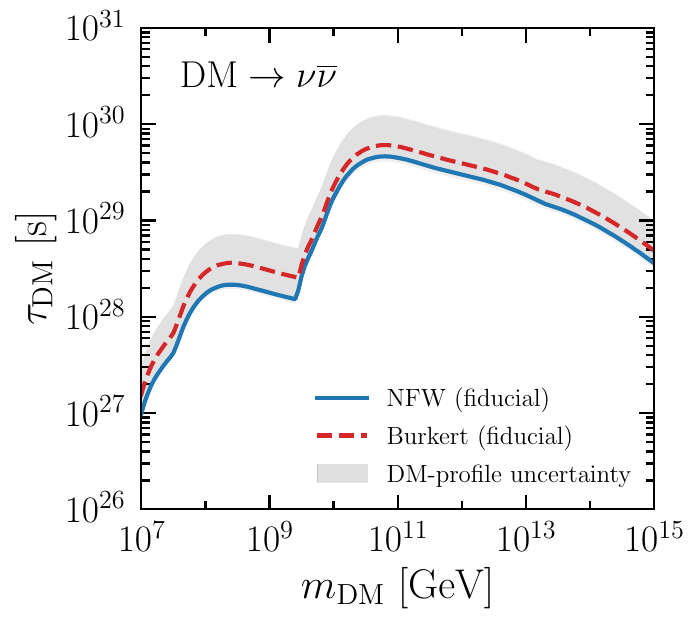}
    \includegraphics[width=0.32\linewidth]{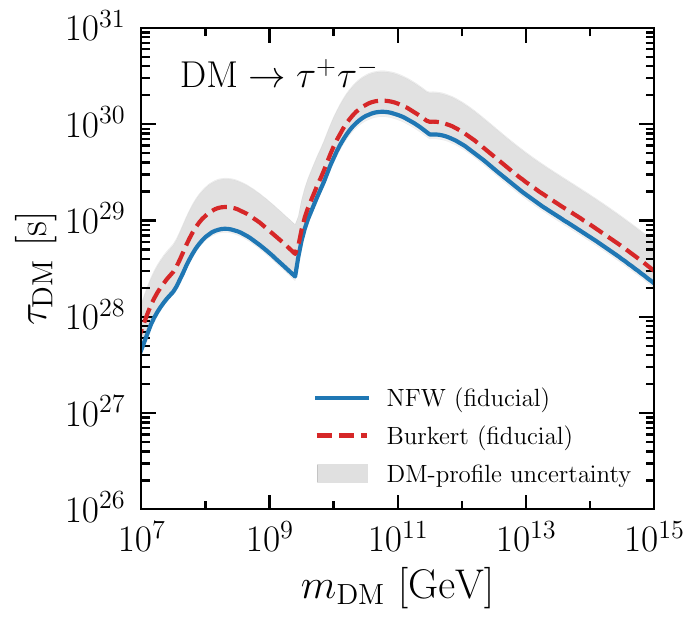}
    \includegraphics[width=0.32\linewidth]{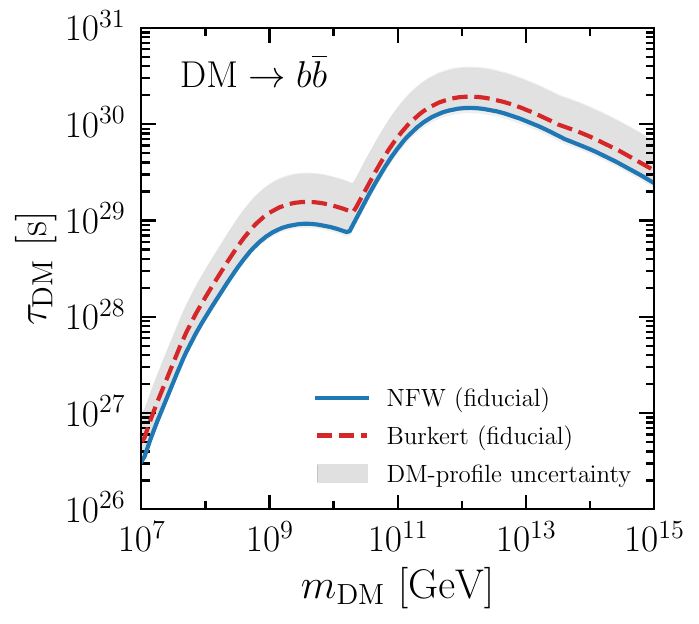}
    \includegraphics[width=0.32\linewidth]{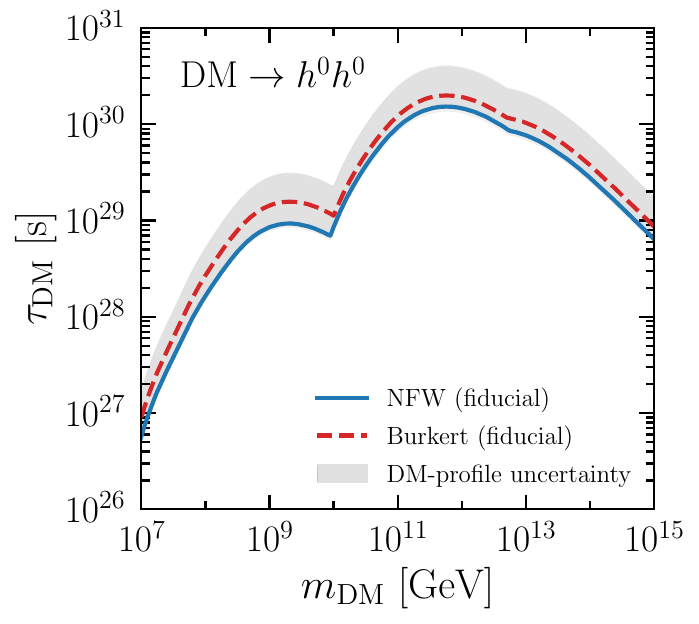}
    \includegraphics[width=0.32\linewidth]{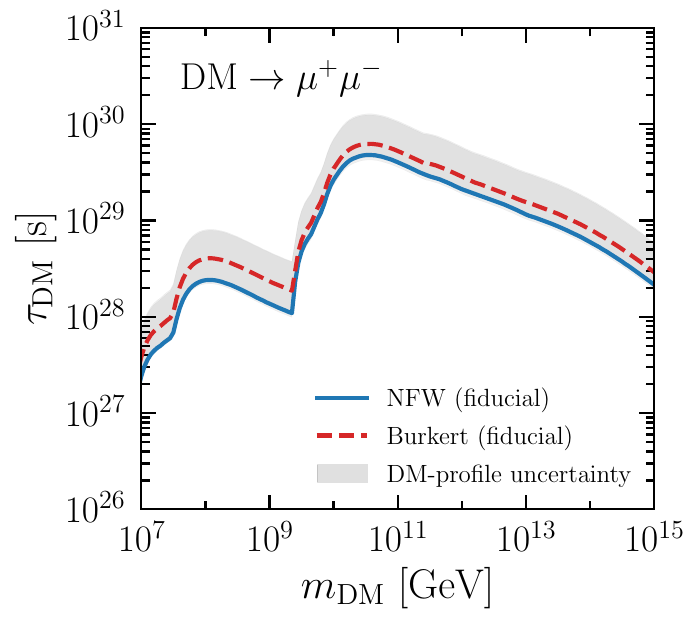}
    \includegraphics[width=0.32\linewidth]{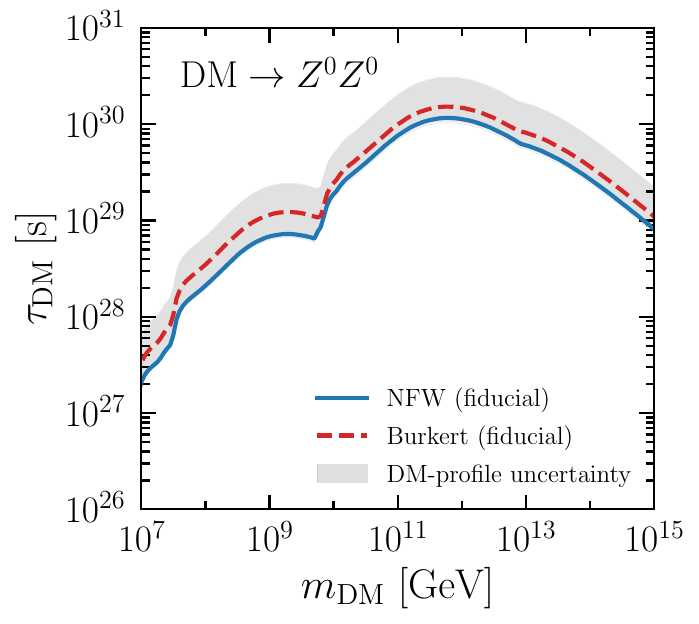}
    \includegraphics[width=0.32\linewidth]{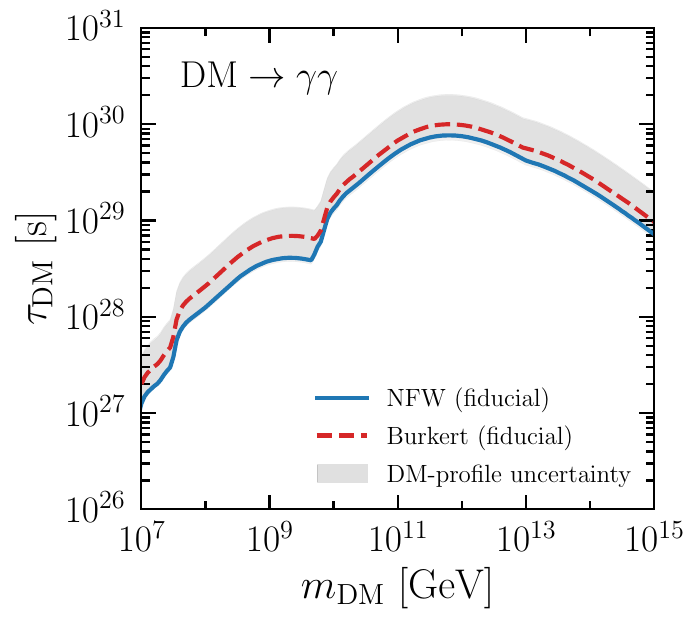}
    \includegraphics[width=0.32\linewidth]{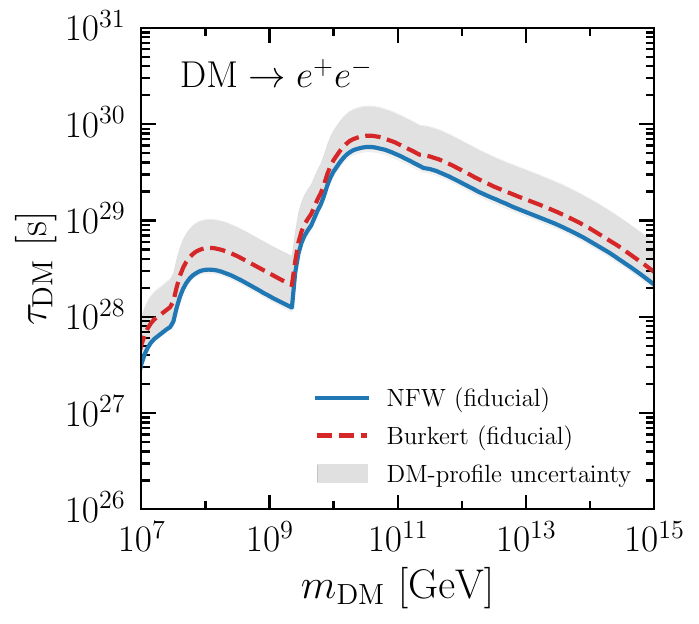}
    \includegraphics[width=0.32\linewidth]{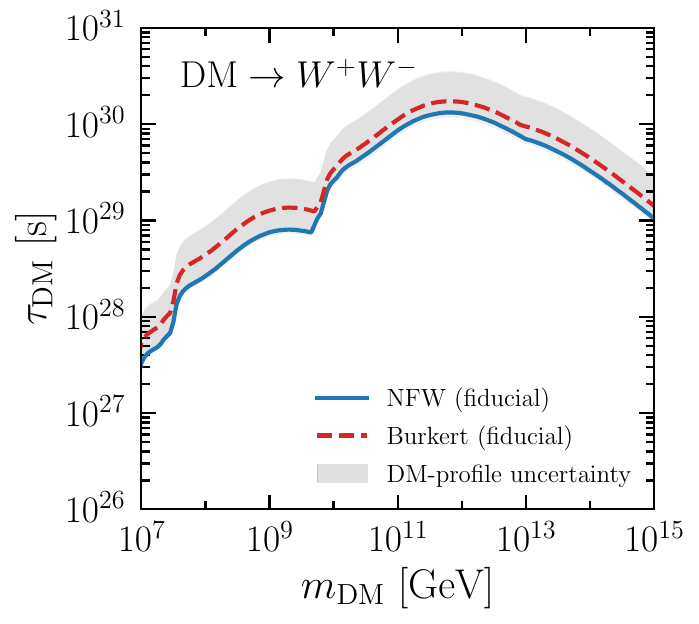}
    \caption{Global lower bounds on the lifetime of decaying SHDM particles at 95\% CL for various decay channels from UHE gamma-ray measurements. The solid blue (dashed red) lines correspond to the NFW (Burkert) density profile with the fiducial values $\rho_\odot = 0.4~{\rm GeV \, cm^{-3}}$ and $R_s = 25~{\rm kpc}$ ($R_c = 25~{\rm kpc}$). The shaded gray region represents the $2\sigma$ uncertainty on the DM lifetime limits from the determination of the galactic DM distribution.}
    \label{fig:channels}
\end{figure}

We report in Fig.~\ref{fig:channels} the constraints obtained in the present work for several decay channels. 

\end{document}